\documentclass[a4paper,twocolumn,nofootinbib]{revtex4-1}

\usepackage{xcolor}
\usepackage{hyperref}
\hypersetup{
	colorlinks,
	linkcolor={red!90!black},
	citecolor={black!10!blue},
	urlcolor={blue!80!black}
}

\usepackage{tikz}
\usepackage{amssymb}
\usepackage{bm,amsmath}
\usepackage{graphicx}

\newcommand{\f}{\frac}

\begin{document}

\title{Theory and simulation for equilibrium glassy dynamics in cellular Potts model of confluent biological tissue}

\author{Souvik Sadhukhan}
\email{ssadhukhan@tifrh.res.in}
\affiliation{TIFR Centre for Interdisciplinary Sciences, Tata Institute of Fundamental Research, Hyderabad - 500046, India}

\author{Saroj Kumar Nandi}
\email{saroj@tifrh.res.in}
\affiliation{TIFR Centre for Interdisciplinary Sciences, Tata Institute of Fundamental Research, Hyderabad - 500046, India}

\begin{abstract}
Glassy dynamics in a confluent monolayer is indispensable in morphogenesis, wound healing, bronchial asthma, and many others; a detailed theoretical framework for such a system is, therefore, important. Vertex model (VM) simulations have provided crucial insights into the dynamics of such systems, but their nonequilibrium nature makes it difficult for theoretical development. Cellular Potts model (CPM) of confluent monolayer provides an alternative model for such systems with a well-defined equilibrium limit. 
We combine numerical simulations of CPM and an analytical study based on one of the most successful theories of equilibrium glass, the random first order transition theory, and develop a comprehensive theoretical framework for a confluent glassy system. We find that the glassy dynamics within CPM is qualitatively similar to that in VM.
Our study elucidates the crucial role of geometric constraints in bringing about two distinct regimes in the dynamics, as the target perimeter $P_0$ is varied. 
The unusual sub-Arrhenius relaxation results from the distinctive interaction potential arising from the perimeter constraint in such systems. Fragility of the system decreases with increasing $P_0$ in the low-$P_0$ regime, whereas the dynamics is independent of $P_0$ in the other regime. The rigidity transition, found in VM, is absent within CPM; this difference seems to come from the nonequilibrium nature of the former.
We show that CPM captures the basic phenomenology of glassy dynamics in a confluent biological system via comparison of our numerical results with existing experiments on different systems.
\end{abstract}

\maketitle
\section{Introduction}
Collective motion of cells in a confluent monolayer is important in morphogenesis \cite{friedl2009,tambe2011,kakkada2018}, cancer metastasis \cite{kakkada2018,streitberger2020}, wound healing \cite{poujade2007,brugues2014,noppe2015,das2015}, bronchial asthma \cite{park2015,atia2018}, 
vertebrate body axis elongation \cite{mongera2018}, and many others. Recent experiments \cite{angelini2011,malinverno2017,palamidessi2019,park2015,mongera2018,garcia2015,kakkada2018} have shown the dynamics in such cellular systems has remarkable similarities with that of a glassy system. 
Glassy dynamics refers to the extreme slowing down, of the order of 12-14 orders of magnitude, with a small change of control parameter without any discernible structural signature or phase transition \cite{debenedetti2001,giulioreview}. The key characteristics of a glassy system, such as the complex stretched exponential relaxation \cite{malinverno2017,park2015,atia2018}, the growing dynamic heterogeneity characterized through higher order susceptibilities \cite{palamidessi2019,park2015,atia2018,angelini2011}, non-Gaussian nature of the displacement distribution \cite{giavazzi2018}, etc, are also displayed in the collective dynamics of cellular systems.
Importance of the problem calls for a detailed theoretical framework for the glassy dynamics in such systems. A confluent monolayer of cells is different from particulate systems in at least two crucial aspects: first, the packing fraction is always unity, and, thus, can not be a control parameter \cite{bi2015,bi2016}, second, the inter-particle interaction potential can be varied as a function of the control parameter.

\begin{figure}
	\centering
	\includegraphics[width=8.6cm]{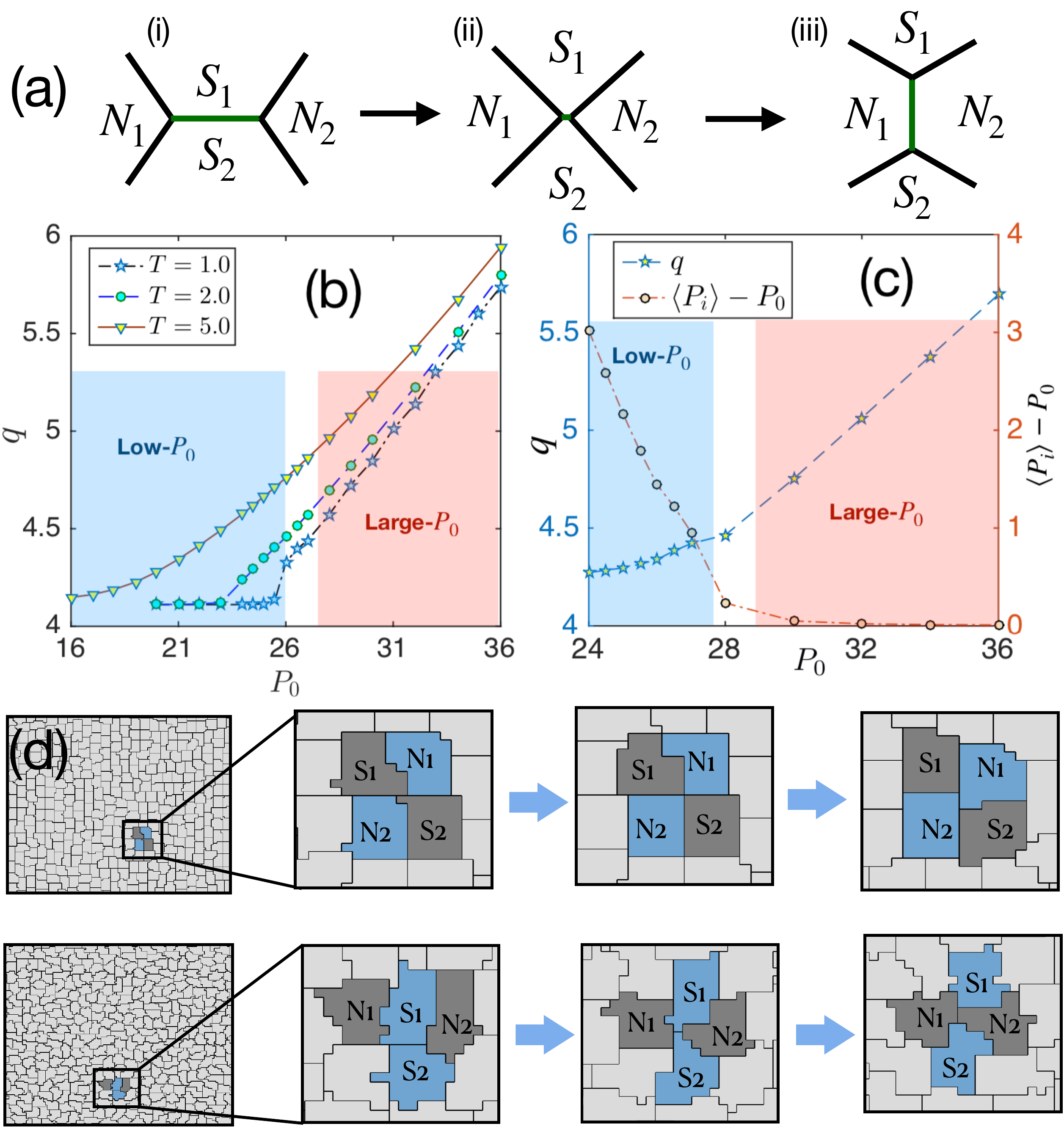}
	\caption{(a) Schematic illustration of $T1$ transition where length of the common edge, in (i), between two cells, $S_1$ and $S_2$, shrinks to zero, in (ii), and finally a common edge between two cells, $N_1$ and $N_2$, that were not neighbors earlier, forms, shown in (iii). This process is represented in vertex models by flipping of an edge, that is directly going from (i) to (iii) when an edge length becomes smaller than a predefined value, leading to nonequilibrium nature of the dynamics.
	(b) Observed shape index, $q$, as a function of $P_0$ at three different $T$; $P_{min}=26$ in our simulations. Lowest value of $q$ is given by geometric restriction in the low-$P_0$ regime and by $P_0$ in the large-$P_0$ regime. (c) $q$ at $T_g$ as a function of $P_0$ tends to saturate in the low-$P_0$ regime and increases linearly with $P_0$ in the large-$P_0$ regime. Right $y$-axis shows $\langle P_i\rangle-P_0$ decreases linearly with increasing $P_0$ in the low-$P_0$ regime and then tends to zero. Each point in (b) and (c) is an average over $10^5$ $t_0$.
	(d) Snapshots of neighbor exchange or a $T1$ transition process in CPM. Upper and lower panels show two $T1$ transition events in systems with $P_0=25$ at $T=2.0$ and $P_0=32$ at $T=0.5$ respectively; we follow the time evolution of four cells shown by the marked regions in the system (left most figures) and show the configurations of these cells at three different times. At the first snapshots, $S_1$ and $S_2$ share a common boundary whereas $N_1$ and $N_2$ don't. The scenario reverses in the last snapshots.}
	\label{msdQoft}
\end{figure}

Inspired by the physics of soap bubbles, vertex models \cite{honda1978,bock2010} that represent individual cells by polygons have provided important insights into the dynamics of such systems \cite{farhadifar2007,staple2010,fletcher2014,bi2015,bi2016,barton2017,sussman2018,krajnc2018,silke2020}. 
Within vertex models, the vertices of the polygons are evolved with certain rules. The cellular perimeter between vertices is either straight by construction or has a constant curvature, whereas in experiments it can deviate arbitrarily from a straight line \cite{mitchel2020,bock2010,atia2018}; how this deviation affects the dynamics remains unknown. An important process governing dynamics in a confluent cellular monolayer is the $T1$ transition or the neighbor exchange process \cite{fletcher2013,fletcher2014}; where an edge between two cells shrink to zero and a new one appears perpendicular to it (see Fig. \ref{msdQoft}(a) for a schematic illustration). Within vertex models, this process is implemented by a perpendicular flip of an edge whenever its length becomes smaller than a predefined value, $d_0$; such an implementation necessarily makes the model nonequilibrium. Moreover, the dynamics crucially depends on $d_0$ \cite{amit2020}, making extension of equilibrium theories for such systems nontrivial and an equilibrium variant of the model is important. Confluent systems have shown to exhibit some unusual glassy properties, understanding the dynamics of such systems should, therefore, also be interesting from the perspective of equilibrium glass transition theories.

The lattice-based cellular Potts models (CPM) \cite{graner1992,glazier1993,hirashima2017} define another important class of models for cellular dynamics and have been applied to single and collective cellular behavior \cite{czirok2013,magno2015,rens2019,kabla2012}, cell sorting \cite{graner1992,glazier1993}, dynamics on patterned surfaces \cite{maree2007}, gradient sensing \cite{camley2017,maree2007}, etc. Despite the widespread applicability of CPM, its glassy aspects remain relatively unexplored. 
To the best of our knowledge, there exists only one such simulation study \cite{chiang2016}, which however did not consider the perimeter constraint and, as we show below, models with and without this constraint are qualitatively different.

The primary difference between CPM and vertex-based models lies in the details of energy minimization \cite{albert2016}. Two crucial aspects of CPM, however, make it advantageous over vertex-based models: it allows arbitrary shape of cell perimeters, and $T1$ transitions are naturally included within CPM. This latter feature allows to study the dynamics of the model in equilibrium, which is the focus of this current study.
Although biological systems are inherently out of equilibrium and activity is crucial, it is important to first understand the behavior of an equilibrium system in the absence of activity, which can be included later \cite{activerfot,nandi2018}. Furthermore, we find that the dynamics in CPM is similar to that in a vertex model and the theoretical framework, developed here, can be applied to the results of vertex-based models \cite{manoj2020}.

The dynamics of CPM in the glassy regime provides an alternative and complementary angle to vertex-models to understand the glassiness in confluent systems. Simulation studies of vertex models have established a rigidity transition that controls the glassy dynamics and the observed shape index (average ratio of perimeter to square root of area) has been interpreted as the structural order parameter of glass transition \cite{park2015,bi2015,bi2016}.
We show that these results are {\em not} generic of confluent systems and, possibly, a consequence of the nonequilibrium nature of the vertex models.
Our aim in this work is twofold: first, we bridge the gap in numerical results through detailed Monte-Carlo (MC) based simulation study of CPM in the glassy regime, and second, we develop the random first order transition (RFOT) theory \cite{kirkthirurmp,lubchenko2007}, one of the most popular theories of glassy dynamics in particulate systems, for a confluent system.

The results of the current work can be summarized as follows: (i) We simulate CPM for a confluent system in glassy regime and find that the qualitative behaviors of the dynamics are similar to those in vertex models. (ii) The target perimeter $P_0$ that parameterizes the interaction potential, plays the role of a control parameter. Geometric restriction brings about two regimes as $P_0$ is varied; dynamics depends on $P_0$ in the low-$P_0$ regime and is independent of $P_0$ in the other, large-$P_0$ regime. (iii) One striking result of our study is the presence of glassy behavior in the large-$P_0$ regime, where vertex models show absence of glassiness, (iv) The rigidity transition of vertex models is absent within CPM; this possibly comes from the difference of how $T1$ transitions are included within the two models. (v) We develop RFOT theory for confluent systems and the theory agrees well with our simulation results. (vi) The perimeter constraint is crucial for the unusual sub-Arrhenius behavior and the system being confluent alone is not sufficient for such behavior. (vii) Velocity distribution is non-Gaussian in the glassy regime in agreement with existing experimental results.
The rest of the paper is organized as follows: We introduce CPM in Sec. \ref{model} and describe some basic characteristics of the system in Sec. \ref{characteristics}. The main results of the work, development of RFOT theory for confluent systems and simulation results in low-$P_0$ and large-$P_0$ regimes are presented in Sec. \ref{results}. We show some critical tests of our theory in Sec. \ref{tests} and comparison with experiments in Sec. \ref{compexpt}. Finally, we conclude the paper in Sec. \ref{disc} via a discussion of our results.

\section{Cellular Potts Model}
\label{model}
The cellular Potts model (CPM), also known as the ``extended large-$q$ Potts model" or the ``Glazier-Graner-Hogeweg (GGH) model" \cite{graner1992,glazier1993,hogeweg2000}, is a lattice based model to simulate the behavior of cellular systems \cite{hirashima2017,maree2007,graner1992}. 
For the CPM in $2D$, we use a square lattice of size $L\times L$ to represent a confluent cell monolayer. Each cell in this lattice consists of a set of lattice sites with the same integer Potts spin ($\sigma$), also known as cell index, where $\sigma \in [0,N]$, $N$ being the total number of cells; $\sigma=0$ is usually reserved for fluid that is absent in our model. The cells in this model are evolved by stochastically updating one lattice site at a time through Monte Carlo (MC) simulation via an effective energy function $\mathcal{H}$ \cite{hirashima2017,albert2016}:
\begin{equation}\label{energyfunction}
\mathcal{H}=\sum_{i=1}^{N} [\lambda_A(A_i-A_0)^2+\lambda_P(P_i-P_0)^2] +J\sum_{\langle kl\rangle} (1-\delta_{\sigma_k,\sigma_l})
\end{equation}
where $\sigma_i, (i \in 1,\ldots,N)$ are cell indices, $N$ is the total number of cells, $A_i$ and $P_i$ are area and perimeter of the $i$th cell, $A_0$ and $P_0$ are target area and target perimeter, chosen to be same for all cells. $\lambda_A$ and $\lambda_P$ are elastic constants related to area and perimeter constraints. The summation in the last term is taken over all nearest neighbor sites $\langle kl\rangle$, $\delta_{\sigma_k,\sigma_l}$ is the Kronecker delta function. $J$ gives the strength of inter-cellular interaction, positive values of $J$ signify repulsion whereas negative $J$ represents attractive interaction.

Cells can be treated as incompressible in $3D$ \cite{jacques2015}. It has been found in experiments that the height of a monolayer remains almost constant \cite{farhadifar2007}. These two findings together allows a $2D$ description of the system with an area constraint leading to the first term in Eq. (\ref{energyfunction}); $A_0$ gives the target cell area and $\lambda_A$ determines the strength of area fluctuation from $A_0$.
On the other hand, mechanical properties of a cell is mostly governed by cellular cortex \cite{jacques2015} and this can be encoded in a perimeter constraint with a target perimeter $P_0$ in the form of the second term in Eq. (\ref{energyfunction}), with $\lambda_P$ determining the strength of perimeter fluctuation. Inter-cellular interactions through different junction proteins like E-Cadherins and effects of pressure, contractility, cell adhesion, etc can be included within an effective interaction term, the third term in Eq. (\ref{energyfunction}).
The last term in $\mathcal{H}$ is proportional to $P_i$ and can be included within the second term with a renormalized value of $P_0$, however, for ease of discussion we keep it separately. 
CPM represents the biological processes for dynamics through an effective temperature $T$
\cite{graner1992,hirashima2017,durand2019}. 
Fragmentation of cells is forbidden \cite{durand2016} in our simulation to minimize noise.
We mainly focus on the model with $J=0$ and get back to the model with $J\neq0$ and $\lambda_P=0$, that was simulated in Ref. \cite{chiang2016}, later in the paper, in Sec. \ref{tests}.

\section{Basic Characteristics of the dynamics}
\label{characteristics}
We next describe some basic characteristics of the dynamics in a confluent system from the perspective of our numerical study of CPM.

{\bf Dynamics is independent of $A_0$:} When total area of the system is fixed, $\mathcal{H}$ in Eq. (\ref{energyfunction}) becomes independent of $A_0$. The change in energy coming from the area term alone for an MC attempt $\sigma_i \to \sigma_j$ between $i$th and $j$th cells is $\Delta\mathcal{H}_{area}=2\lambda_A(1-A_i+A_j)$ that is independent of $A_0$. 
Since $A_0$ dependence of dynamics can only come through $\Delta\mathcal{H}_{area}$, the dynamics becomes independent of $A_0$. This argument can also be extended for a polydisperse system. The input shape index, $s_0=P_0/\sqrt{A_0}$, therefore, cannot be a control parameter for the dynamics and should be viewed as a dimensionless perimeter; this result was also found for voronoi model dynamics \cite{yang2017}. $P_0$, on the other hand, parameterizes the interaction potential and plays the role of a control parameter.

{\bf Two different regimes of $P_0$:} 
The observed shape index, $q=\langle P_i/\sqrt{A_i}\rangle$, where $\langle\ldots\rangle$ denotes average over all cells, tends to a constant with decreasing $P_0$ (Fig. \ref{msdQoft}b). $q$ seems to be the structural order parameter of glass transition in vertex models \cite{park2015,bi2015,bi2016}, however, as we show below, such an interpretation is {\em not} applicable for CPM.
$P_i$ for a fixed $A_i$ has a minimum value, $P_{min}$, that depends on geometric constraints, here confluency and underlying lattice. 
When $P_0$ is below $P_{min}$, $P_i$ of most cells cannot satisfy the perimeter constraint in Eq. (\ref{energyfunction}) as they remain stuck around $P_{min}$. 
At high $T$ fluid regime, when dynamics is fast, cell boundaries are irregular leading to larger values of $P_i$ and $q$; 
but, at low $T$ glassy regime, when dynamics is slow, cell boundaries tend to be regular leading to lower values of $P_i$ and $q$. 
Figure \ref{msdQoft}(b) shows $q$ at three different $T$ as a function of $P_0$; $q$ decreases with decreasing $P_0$ at a fixed $T$ and saturates to $4.11$ (the quantitative value is lattice-dependent, however, the qualitative behavior, we expect, to be independent of the lattice). 
The lowest value of $q$ is dictated by the geometric restriction in the low-$P_0$ regime. 
Our interpretation is consistent with the simulation results in voronoi models \cite{sussman2018,sussmansoftmatter2018} as well as the fact that $q$ in a large class of distinctly different systems has similar values \cite{li2020,atia2018}.

On the other hand, when $P_0>P_{min}$, the large-$P_0$ regime, most cells are able to satisfy the perimeter constraint and the lowest value of $q$ is governed by $P_0$, as deviation from $P_0$ costs energy. We show $q$ at the glass transition temperature, $T_g$, (defined as the $T$ when relaxation time becomes $10^4$) as a function of $P_0$ in Fig. \ref{msdQoft}(c); $q$ tends to a constant in the low-$P_0$ regime whereas it increases linearly with $P_0$ in the large-$P_0$ regime. The geometric restriction is clearer in the plot of $\langle P_i\rangle-P_0$; it decreases linearly with increasing $P_0$ in the low-$P_0$ regime and then tends to zero. 
The interfacial tension, defined as $\gamma=\partial \mathcal{H}/\partial P_i\propto(P_i-P_0)$ \cite{magno2015}, is non-zero along cell boundaries in the low-$P_0$ regime and becomes zero in the large-$P_0$ regime as $P_0$ increases.

{\bf $T1$ transitions within CPM:} Dynamics in a biological tissue proceeds via a series of complicated biochemical processes that are simply represented via an effective temperature $T$ within CPM \cite{hirashima2017,durand2019}.
This is an extreme level of simplification from biological perspective, however, is convenient from theoretical aspects. 
At the coarse grained level, $T1$ transitions where cells exchange their neighbors \cite{fletcher2014} are crucial for dynamics in a confluent system. As discussed in the introduction, and illustrated in Fig. \ref{msdQoft}(a), implementation of $T1$ transition within vertex models is nonequilibrium in nature.
On the other hand, $T1$ transitions are naturally included within CPM. We show two such $T1$ transition processes from our simulations in Fig. (\ref{msdQoft}(d)) for $P_0=25$ at $T=2.0$ (upper panel) and for $P_0=32$ at $T=0.5$ (lower panel). The $T1$ transitions within CPM are equilibrium processes and their rates depend on $T$ and $P_0$; this crucial difference, compared to vertex models, is important from theoretical perspective as it allows a well-defined equilibrium limit of the model and makes it easier to extend equilibrium theories of glassy dynamics for confluent systems.
More important, as discussed in Sec. \ref{results}, this difference of how $T1$ transitions are implemented is, possibly, related to the absence of glassy dynamics in the large-$P_0$ regime as well as the identification of $q$ as the structural order parameter of glassy dynamics within the vertex models.

%; a cell boundary (shown in (i) in Fig. \ref{msdQoft}(a)) between two cells, $S_1$ and $S_2$, shrinks to zero (shown in (ii)) and a new cell boundary forms between two other cells, $N_1$ and $N_2$, that were initially not sharing common boundary as in (iii) in Fig. \ref{msdQoft}(a). Within vertex model, $T1$ transitions are implemented via a perpendicular flip of an edge (from step (i) directly to (iii)), when its length becomes smaller than a predefined value $d_0$. The dynamics crucially depends on the choice of $d_0$ and the process necessarily makes the dynamics nonequilibrium and extension of equilibrium theories complicated.

\begin{figure*}
	\centering
	\includegraphics[width=0.7\textwidth]{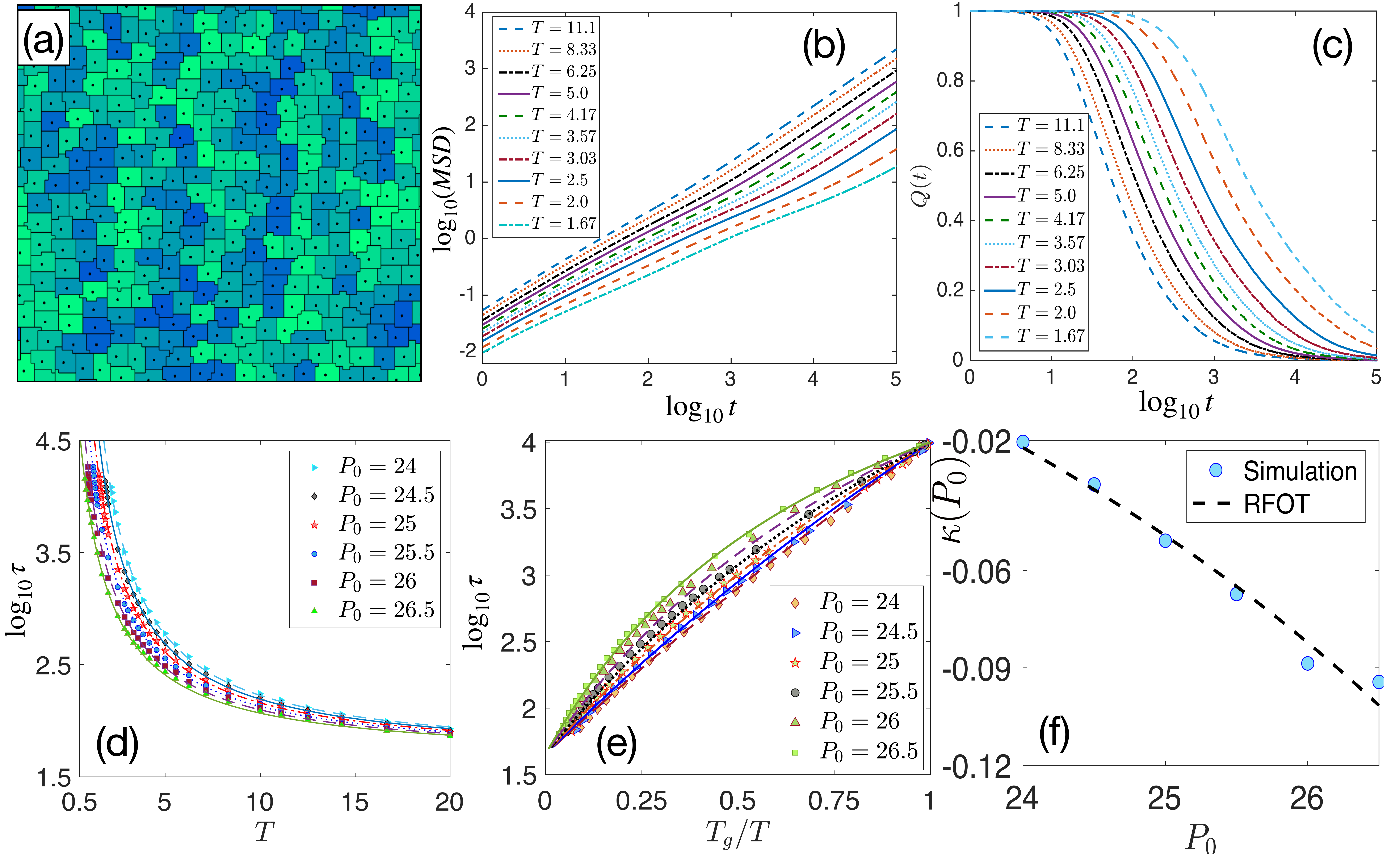}
	\caption{Behavior of CPM in the low-$P_0$ regime. (a) Typical configuration of a system at $P_0=25$ and $T=2.5$, close to $T_g$. Due to the underlying lattice structure, minimum perimeter configuration for a certain area is a square that shows up in the low $T$ configuration. 
		(b) Mean square displacement (MSD) and (c) self-overlap function, $Q(t)$, as a function of time $t$ for $P_0=25$ show typical glassy behaviors where growth of MSD and decay of $Q(t)$ become slower with decreasing $T$. (d) Relaxation time $\tau$ as a function of $T$ for different $P_0$, symbols are simulation data and lines are the corresponding RFOT theory plots (Eq. \ref{rfotlowP0}). (e) Angell plot in this regime shows sub-Arrhenius relaxation, symbols are data and lines are RFOT theory predictions. (f) Simulation data (symbols) for kinetic fragility, $\kappa(P_0)$, in this regime also agree well with the RFOT theory prediction (line).}
	\label{lowP0regime}
\end{figure*}

\section{Results}
\label{results}
We now present our theory for the glassy dynamics in a confluent monolayer. The simulation results for glassy dynamics within CPM, both in the low-$P_0$ and large-$P_0$ regimes, are presented along with the theory.

{\bf RFOT theory for CPM:} The physics of glassy dynamics, even for particulate systems in equilibrium, continues to be debated leading to many different theories of glass transition \cite{giulioreview,hecksher2015}. One of the most successful theories is the random first order transition (RFOT) theory due to Wolynes, Kirkpatrick and Thirumalai \cite{kirk1989,kirkthirurmp,lubchenko2007,bouchaud2004,parisi2010}. Despite the intricate microscopic phenomenology, the theory leads to a simple set of predictions that agree well with experiments on wide set of glassy systems \cite{lubchenko2007}; our goal in this work is to develop this theory for a confluent system to understand the effect of $P_0$ on the glassy dynamics.

Within RFOT theory, a glassy system consists of mosaics of different states; a nucleation-like argument gives the typical length scale of these mosaics \cite{lubchenko2007}. Consider a region of length scale $R$ in dimension $d$, the energy cost for rearrangement (changing state) of this region is
\begin{equation}\label{rfotenergySM}
\Delta F=-f \Omega_d R^d+\Gamma S_d R^\theta,
\end{equation}
where $f$ is the decrease in energy per unit volume due to the rearrangement, $\Omega_d$ and $S_d$, volume and surface of a unit hypersphere, $\Gamma$, the surface energy cost per unit area and $\theta \leq(d-1)$ is the exponent relating surface area and length scale of a region. Within RFOT theory, the drive to reconfiguration is entropic in nature and given by the configurational entropy, $s_c$, that can be thought of as the difference of total entropy of the system and its entropy if it was allowed to crystallize. Thus, $f=k_BT s_c$, where $k_B$ is the Boltzmann constant. Minimizing Eq. (\ref{rfotenergySM}) with respect to $R$, we get the typical length scale, $\xi$, for the mosaics as
\begin{equation}\label{xieqSM1}
\xi=\left(\f{\theta S_d \Gamma}{d\Omega_dk_BTs_c}\right)^{1/(d-\theta)}.
\end{equation}

In general, the interaction potential, $\Phi$, of the system determines both $s_c$ and $\Gamma$. Within CPM, the interaction potential is parameterized through $P_0$, thus, $\Phi=\Phi(P_0)$. The temperature dependence of $\Gamma$ is assumed to be linear \cite{wolynesbook}, thus, $\Gamma=\Xi[\Phi(P_0)]T$ and we write Eq. (\ref{xieqSM1}) as
\begin{equation}
\xi=\left(\f{D\Xi[\Phi(P_0)]}{s_c[\Phi(P_0)]}\right)^{1/(d-\theta)},
\end{equation}
where $D=\theta S_d/dk_B\Omega_d$ is a constant. Within RFOT theory, relaxation dynamics of the system refers to relaxations of individual mosaics. The energy barrier associated for the relaxation of a region of length scale $\xi$ is $\Delta(\xi)=\Delta_0\xi^\psi$, where $\Delta_0$ is an energy scale and $\psi$ is an exponent. The relaxation time then becomes  $\tau=\tau_0\exp({\Delta_0\xi^\psi}/{k_BT})$, where $\tau_0$ is a microscopic time scale independent of $T$, but can depend on interatomic interaction potential, hence, on $P_0$. Taking $\Delta_0=\kappa T$, where $\kappa$ is a constant \cite{lubchenko2007,kirkthirurmp} and setting $k_B$ to unity, we obtain $\tau$ as
\begin{equation}\label{taueq1}
\ln\left(\f{\tau}{\tau_0}\right)=\kappa\left\{\f{D\Xi[\Phi(P_0)]}{s_c[\Phi(P_0)]}\right\}^{\psi/(d-\theta)}.
\end{equation}
Following Refs. \cite{kirk1989,kirkthirurmp} we take $\theta=\psi=d/2$ and then Eq. (\ref{taueq1}) can be written as
\begin{equation}\label{rfotreltimebasic}
\ln\left(\f{\tau}{\tau_0}\right)=\f{E\Xi[\Phi(P_0)]}{s_c[\Phi(P_0)]},
\end{equation}
where $E=\kappa D$ is another constant. The theory presented here is similar in spirit with that for a network material obtained by Wang and Wolynes \cite{wang2013}. Eq. (\ref{rfotreltimebasic}) gives the general form of RFOT theory for the CPM; we obtain the detailed forms of $\Xi(P_0)$ and $s_c(P_0)$ for different systems and regimes that we consider below.
Our approach is perturbative in nature and we look at the effect of $P_0$ by expanding the potential around a reference state.

{\bf Low $P_0$ regime:} As discussed above, $P_i$ for most cells are less than $P_0$ in this regime. 
Figure \ref{lowP0regime}(a) shows a typical configuration of cells and their centers of mass for a system, close to glass transition. The mean-square displacement ($MSD$) and the self-overlap function, $Q(t)$, (defined in Appendix \ref{simdet}) as a function of time $t$ show typical glassy behavior (Figs. \ref{lowP0regime}b,c).
We define relaxation time, $\tau$, as $Q(t=\tau) = 0.3$. 
%For a particular $P_0$, the glass transition temperature, $T_g$, is defined as $\tau(T_g)=10^4$. 

We now develop the RFOT theory for CPM in this regime, where cells are not able to satisfy the perimeter constraint and the dynamics depends on $P_0$. Within our perturbative approach we treat a confluent system with $P_0=P_0^{ref}$ as our reference system around which we expand the effect of varying $P_0$ on $s_c$ and $\Xi$. Thus, we have
\begin{align}\label{scXiexpansionSM}
s_c[\Phi(P_0)]&=s_c[\Phi(P_0^{ref})]  +\f{\delta s_c[\Phi(P_0)]}{\delta\Phi(P_0)}\bigg|_{P_0^{ref}}\delta\Phi(\delta P_0) +\ldots \nonumber\\
\Xi[\Phi(P_0)]&=\Xi[\Phi(P_0^{ref})] +\frac{\delta \Xi[\Phi(P_0)]}{\delta \Phi(P_0)}\bigg|_{P_0^{ref}}\delta\Phi(\delta P_0)+\ldots
\end{align}
where $\delta P_0=(P_0-P_0^{ref})$ and we have ignored higher order terms. Within RFOT theory, glassiness, that is the abrupt slowing down of dynamics at low $T$, results from a thermodynamic transition taking place at an even lower $T$, known as the Kauzmann temperature \cite{kauzmann1948}, $T_K$, where the configurational entropy of the system vanishes and $\tau$ diverges.  Thus, $s_c[\Phi(P_0^{ref})]$ can be written as 
\begin{equation}\label{refsc_SM}
s_c[\Phi(P_0^{ref})]=\Delta C_p(T-T_K)/T_K,
\end{equation}
where $\Delta C_p$ is the difference of specific heats of the liquid and the periodic crystalline phase.
Within linear order, $\delta\Phi(P_0-P_0^{ref})$, the change in potential due to a variation in $P_0$ from the reference state, can be taken to be proportional to $(P_0-P_0^{ref})$:
\begin{align}\label{scXieq2SM}
&\f{\delta s_c[\Phi(P_0)]}{\delta\Phi(P_0)}\bigg|_{P_0=P_0^{ref}}\delta\Phi(P_0-P_0^{ref})=\bar{\varkappa}_c(P_0-P_0^{ref}) \nonumber\\
&\frac{\delta \Xi[\Phi(P_0)]}{\delta \Phi(P_0)}\bigg|_{P_0=P_0^{ref}}\delta\Phi(P_0-P_0^{ref})=-\bar{\varkappa}_s(P_0-P_0^{ref}).
\end{align}
Using Eqs. (\ref{scXiexpansionSM}-\ref{scXieq2SM}) in Eq. (\ref{rfotreltimebasic}), we obtain
\begin{equation}\label{rfotlowP0SM}
\ln\left(\f{\tau}{\tau_0}\right)=\f{k_1-k_2(P_0-P_0^{ref})}{T-T_K+\varkappa_c(P_0-P_0^{ref})}
\end{equation}
where $k_1=T_KE\Xi[\Phi(P_0^{ref})]/\Delta C_p$, $k_2=T_KE\bar{\varkappa}_s/\Delta C_p$ and $\varkappa_c=T_K\bar{\varkappa}_c/\Delta C_p$ are all constants. The value of $P_0^{ref}$ depends on the average cell area, for the results presented in this work, the average cell area is 40 and we find $P_0^{ref}=23$ provides a good description of the data. Thus, using $P_0^{ref}=23$, we obtain
\begin{equation}\label{rfotlowP0}
\ln\left(\f{\tau}{\tau_0}\right)=\f{k_1-k_2(P_0-23)}{T-T_K+\varkappa_c (P_0-23)}.
\end{equation}
The constants $k_1$, $k_2$, $T_K$ and $\varkappa_c$ are independent of $T$ and $P_0$; they only depend on the microscopic details of a system and dimension. For a given system, we treat these constants as fitting parameters in the theory and obtain their values from fit with simulation data. Note that $\tau_0$ depends on the high $T$ properties of the system, which is nontrivial and will be explored elsewhere. Our analysis in the low-$P_0$ regime shows that $P_0$-dependence of $\tau_0$ is weaker and can be taken as a constant.

\begin{figure*}
	\includegraphics[width=15cm]{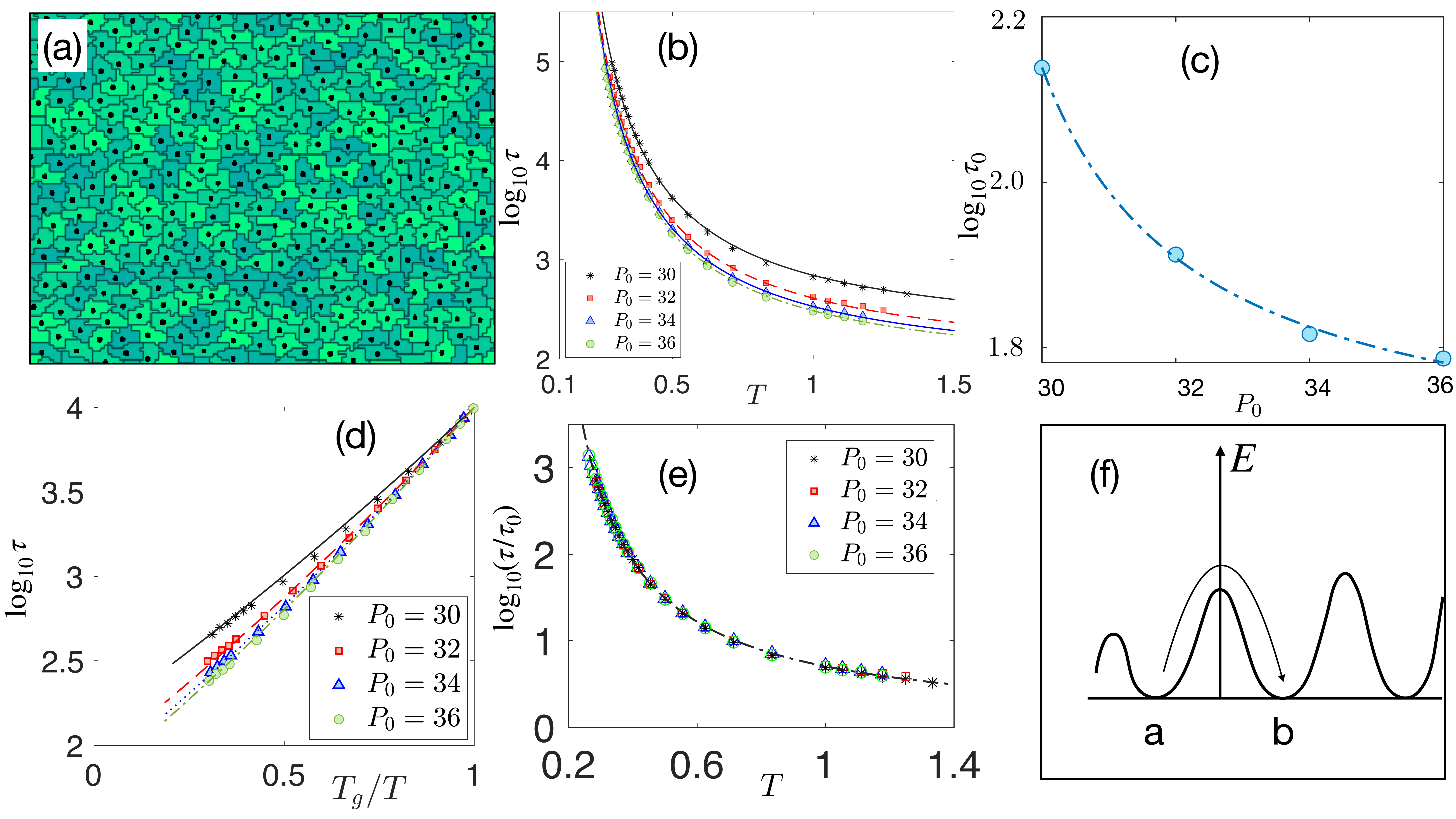}
	\caption{Behavior of CPM in the large-$P_0$ regime. (a) Typical configuration of the system with $P_0=34$ and $T=0.5$ close to $T_g$. The cell boundaries become irregular to satisfy the perimeter constraint. (b) Relaxation time $\tau$ as a function of $T$, symbols are data and lines are RFOT theory fits (Eq. \ref{largeP0rfot}). (c) $\tau_0$ as a function of $P_0$. The line is a fit with a function $\ln\tau_0(P_0)=a+b/(P_0-c)$ with $a=3.79$, $b=2.58$ and $c=27.72$. (d) Angell plot for the same data (symbols) as in (b), lines are RFOT theory plots (Eq. \ref{largeP0rfot}). (e) $\tau/\tau_0$ for different $P_0$ follow a master curve, the line is RFOT theory result. The data collapse illustrates that glassiness in this regime is independent of $P_0$. (f) Schematic illustration of the dynamics as a barrier crossing between two equal-energy minima.}
	\label{largeP0plot}
\end{figure*}

The minimum possible perimeter in our simulation is 26 (Appendix \ref{simdet}) and we expect the critical $P_0$ separating the two regimes to be somewhere between 27 and 28. We first concentrate on the results for $P_0=24$ to $26.5$ and present $\tau$ as a function of $T$ for different $P_0$ in Fig. \ref{lowP0regime}(d).
We fit one set of data presented in Fig. \ref{lowP0regime}(d) with Eq. (\ref{rfotlowP0}) and obtain the parameters as follows: $\tau_0=45.13$, $k_1=14.78$, $k_2=1.21$, $T_K=0.0057$ and $\varkappa_c=0.31$.
Note that with these constants fixed, there is {\em no other fitting parameter} in the theory, we now show the plot of Eq. (\ref{rfotlowP0}), as a function of $T$ for different values of $P_0$ with lines in Fig. \ref{lowP0regime}(d). Figure \ref{lowP0regime}(e) shows the same data in Angell plot representation that shows $\tau$ as a function of $T_g/T$ in semi-log scale. All the curves meet at $T=T_g$ by definition. The simulation data agree well with RFOT predictions at low $T$ where the theory is applicable.

When $\tau\sim \exp[C_A/T]$, where $C_A$ is a constant, we obtain a straight line in the Angell plot representation of $\tau$, as in Fig. \ref{lowP0regime}(e); this is the well-known Arrhenius behavior \cite{giulioreview,debenedetti2001}. Super-Arrhenius behavior, where $\tau$ changes faster than the Arrhenius law, leads to the relaxation time curves below this straight line whereas sub-Arrhenius behavior, that is slower than the Arrhenius law, shows up as the curves being above this line in Angell plot representation.
In most equilibrium glassy systems, $\tau$ increases similar to or faster than Arrhenius law \cite{lubchenko2007,giulioreview,debenedetti2001}.
One striking feature of the Angell plot in Fig. \ref{lowP0regime}(e) is the sub-Arrhenius nature of $\tau$. 
 Similar results were reported for voronoi and vertex models in Refs. \cite{sussman2018,li2021} demonstrating similarities between CPM and vertex-based models. 
Within our RFOT theory, the sub-Arrhenius relaxation appears due to the distinctive interaction potential imposed by the perimeter constraint, in a regime controlled by geometric restriction, and appears when the system is about to satisfy the perimeter constraint. An important characteristic of this regime is that $[T_K-\varkappa_c (P_0-23)]$ becomes negative. We get back to this point in Sec. \ref{tests} when we subject our RFOT theory to more stringent tests.

One can define a kinetic fragility, $\kappa(P_0)$, and fit the simulation data for different $P_0$ with the form $\ln(\tau/\tau_0)=1/(\kappa(P_0)[T/T_K^{eff}-1])$. We present $\kappa(P_0)$ in Fig. \ref{lowP0regime}(f) where symbols are values obtained from fits with simulation data and the dotted line is theoretical prediction, the agreement, again, is remarkable. Fragility of the system decreases as $P_0$ increases and $\kappa(P_0)$ becomes more negative consistent with stronger sub-Arrhenius behavior.

{\bf Large-$P_0$ regime:} In this regime most cells satisfy the perimeter constraint, cell boundaries are nonlinear [Fig. \ref{largeP0plot}(a)], and dynamics becomes independent of $P_0$ implying constant values of $\Xi$ and $s_c$. Then the RFOT theory, Eq. (\ref{rfotreltimebasic}) after a straightforward algebra, becomes
\begin{equation}\label{largeP0rfot}
\ln\left(\f{\tau}{\tau_0(P_0)}\right)=\f{\Xi}{T-T_K}.
\end{equation}
Although $P_i=P_0$ on the average, there are fluctuations of $P_i$ around $P_0$ when $T\neq0$. The interaction potential is governed by these fluctuations that are stronger at higher $T$. Thus, $P_0$-dependence of $\tau_0$ is important in this regime.

\begin{figure*}
	\centering
	\includegraphics[width=14cm]{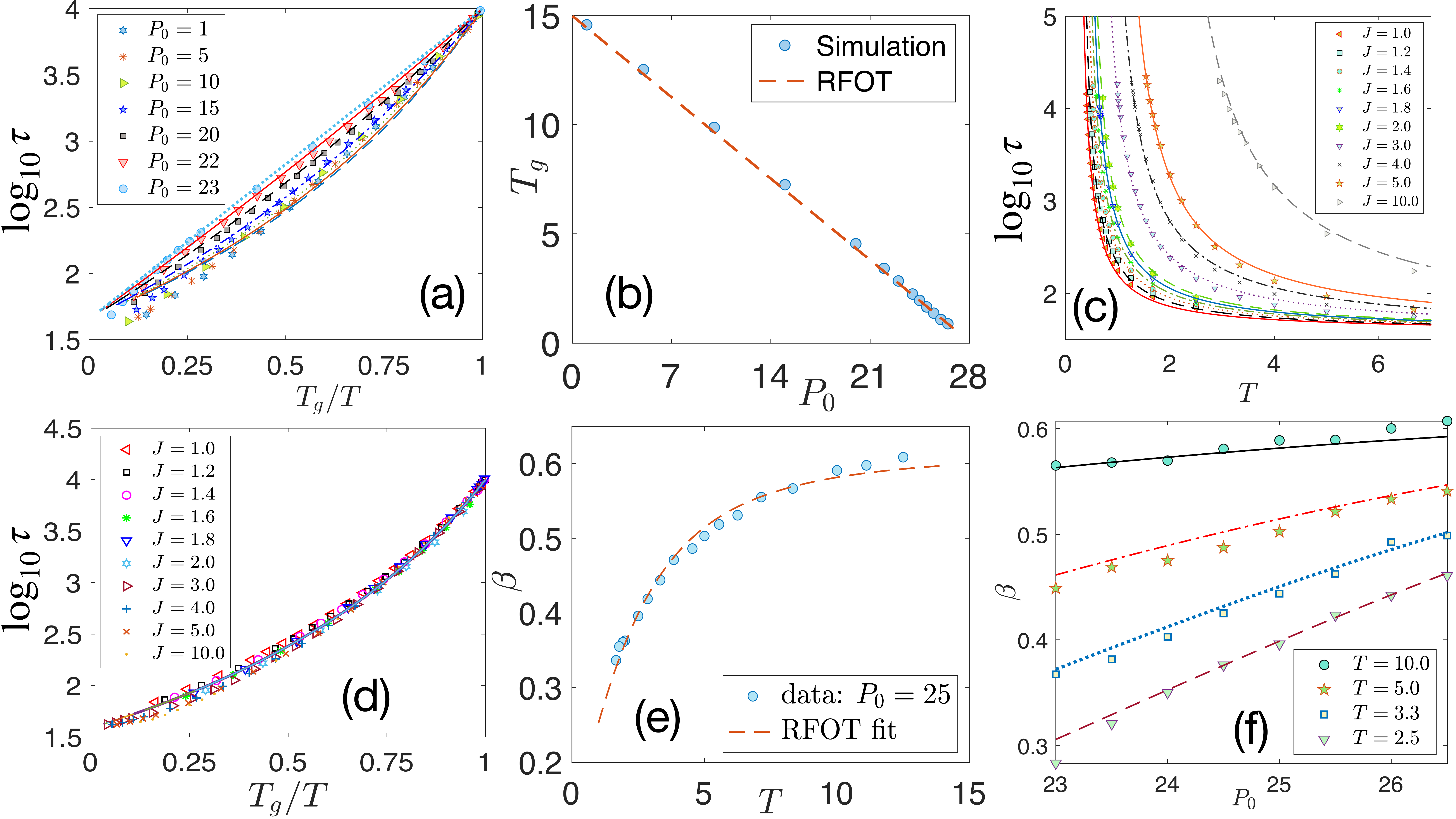}
	\caption{Tests of our extended RFOT theory. (a) Theory predicts super-Arrhenius behavior for $P_0\leq23$. Angell plot for low-$P_0$ simulation data (symbols) agree well with the RFOT theory predictions, Eq. (\ref{rfotlowP0}) (lines). (b) Comparison of $T_g$ at different $P_0$ between simulation (symbols) and RFOT theory (dashed line). (c) $\tau$ for the model with $\lambda_P=0$ and different values of $J$, symbols are simulation data and lines are RFOT theory (Eq. \ref{rfotchiangmodelSM}). (d) Simulation data (symbols) for this system show super-Arrhenius behavior and constant fragility and agree well with the RFOT theory (lines). (e) Stretching exponent $\beta$ for $P_0=25$ as a function of $T$. Fit of simulation data with the RFOT theory expression, Eq. (\ref{rfotbeta}), gives $\mathcal{A}=0.62$ and $\mathcal{B}=0.3$. (f) Trends of $\beta$ as a function of $P_0$ at different $T$ agree well with the RFOT theory predictions, Eq. (\ref{rfotbeta}), with $\mathcal{A}$ and $\mathcal{B}$ obtained from the fit in (e).}
	\label{summarycomp}
\end{figure*}

Figure \ref{largeP0plot}(b) shows $\tau$ as a function of $T$; they clearly vary for different $P_0$ and this difference comes from $P_0$-dependence of $\tau_0$. We fit Eq. (\ref{largeP0rfot}) with one set of data and obtain $\Xi=1.54$, $T_K=0.052$ and a corresponding value for $\tau_0(P_0)$. Keeping $\Xi$ and $T_K$ fixed, we next fit rest of the data to obtain $\tau_0(P_0)$. The fits are shown by lines in Fig. \ref{largeP0plot}(b) and $\tau_0(P_0)$ is shown in Fig. \ref{largeP0plot}(c) where the line is a proposed form: $\ln\tau_0(P_0)\sim 1/(P_0-\text{constant})$. 
Figure \ref{largeP0plot}(d) shows the Angell plot representation of the same data as in (b) and lines are the corresponding RFOT theory plots. 
Figure \ref{largeP0plot}(e) shows $\tau/\tau_0$ as a function of $T$ for different values of $P_0$, all the data following a master curve support our hypothesis that $P_0$-dependence in this regime comes from $\tau_0(P_0)$.

More important, one would expect no glassy behavior in this regime if a rigidity transition, as in the vertex model
\cite{bi2015,bi2016}, controlled glassiness . In contrast, CPM shows the presence of glassy behavior even in this regime, where $q$ at $T_g$ is proportional to $P_0$ [Fig. \ref{msdQoft}(c)], thus, $q$ cannot be an order parameter for the glass transition in CPM.
As apparent from Fig. \ref{largeP0plot}(a), the configuration in this regime is disordered; at strictly zero $T$, the minimum energy of the system is zero as cells are able to satisfy the area and perimeter constraints. However, the ground state is degenerate with a large multiplicity \cite{staple2010}. Dynamics in this regime can be viewed as exploration of the system among these equal energy ground state configurations. Consider two such states, shown by $a$ and $b$ in a schematic energy landscape plot \cite{debenedetti2001} in Fig. \ref{largeP0plot}(f). The energy difference between the states is zero, but they are separated by a barrier; any dynamics necessarily requires change in area, even for the moves where perimeter can be kept constant, leading to a barrier. 
Within vertex model, the nonequilibrium implementation of $T1$ transitions may allow transition between two such states (going from (i) to (iii) in Fig. \ref{msdQoft}(a)) without a cost; this is, {\em possibly}, the source of dynamics even at zero $T$, leading to the rigidity transition. However, the absence of this nonequilibrium process within CPM forbids any dynamics at strictly zero $T$ as barrier crossings are not allowed; this rules out the rigidity transition of vertex models \cite{bi2015} in CPM. In contrast to the vertex models, this rigidity transition is also absent in equilibrium voronoi models \cite{sussman2018,sussmansoftmatter2018}, the source of this difference remains unclear. A more detailed understanding of the effect of the $T1$ transitions in vertex models is outside the scope of the current work.

\section{Further tests of extended RFOT theory} 
\label{tests}
Having demonstrated that our RFOT theory captures the key characteristics of glassiness in a confluent system, we now subject our theory to stringent tests through three different questions:

Within the theory sub-Arrhenius behavior is found when the effective $T_K$ is negative, i.e., $T_K-\varkappa_c (P_0-23)<0$ (Eq. (\ref{rfotlowP0})). This implies super-Arrhenius behavior for $P_0\leq (T_K+23 \varkappa_c)/\varkappa_c\approx23$. We now simulate the system in this regime and show the Angell plot in Fig. \ref{summarycomp}(a) where symbols represent simulation data and the corresponding lines are RFOT theory predictions. We emphasize that these curves are {\em not} fits, we simply plot Eq. (\ref{rfotlowP0}) with the constants as obtained earlier. All the relaxation curves for different $P_0$ are super-Arrhenius as predicted by the theory. Ref. \cite{li2021} show super-Arrhenius behavior in voronoi model simulations in the low-$P_0$ regime, consistent with our theory.
Figure \ref{summarycomp}(b) shows comparison of $T_g$, obtained from simulation and RFOT theory, for different $P_0$. 

Next, we illustrate that the sub-Arrhenius behavior and negative kinetic fragility is a result of the perimeter constraint in Eq. (\ref{energyfunction}) and confluency alone is not enough for such behavior. We simulate a confluent system with $\lambda_P=0$ and study the glassy behavior as a function of $J$ (Eq. \ref{energyfunction}).
%: the system not only shows super-Arrhenius behavior, but also a constant fragility (i.e., a master curve in Angell plot). 
%This result, a posteriori can be associated with the absence of any special constraint when $\lambda_P=0$.
%The extended RFOT theory for this system also captures the results. 
%Since the repulsive interaction only comes in the form of $J$, the trends of variation of $s_c$ and $\Xi$ with increasing $J$ become opposite to that in the low-$P_0$ regime above. 
Considering the reference system at a moderate value of $J$, a similar calculation as above gives the RFOT expression for $\tau$ as
\begin{equation}\label{rfotchiangmodelSM}
\ln\left(\f{\tau}{\tau_0}\right)=\f{k_1+k_2J}{T-T_K-\varkappa_c J},
\end{equation}
where $k_1$, $k_2$, $T_K$ and $\varkappa_c$ are constants. 
%Note the opposite signs of the constants $k_2$ and $\varkappa_c$ in Eq. (\ref{rfotchiangmodelSM}) and Eq. (\ref{rfotlowP0SM}). 
Fitting Eq. (\ref{rfotchiangmodelSM}) with simulation data for $J=1$, we obtain $k_1=0.284$, $k_2=0.84$, $T_K=0.04856$ and $\varkappa_c=0.157$. %Comparison of Eq. (\ref{rfotchiangmodelSM}) with the simulation data is presented in the main text, Figs. 4(c) and (d).
Figure \ref{summarycomp}(c) shows simulation data for $\tau$ as a function of $T$ for different $J$ (symbols) as well as the corresponding RFOT theory (Eq. \ref{rfotchiangmodelSM}) predictions (lines). The Angell plot corresponding to these data are shown in Fig. \ref{summarycomp}(d). The system exhibits super-Arrhenius relaxation and data for different $J$ follow a master curve, in agreement with theory. These results are important from at least two aspects: first, they show that systems with and without the perimeter constraint are qualitatively different \cite{chiang2016}, and second, that the presence of the perimeter constraint is crucial for the sub-Arrhenius behavior of the system.

\begin{figure*}
	\centering
	\includegraphics[width=14cm]{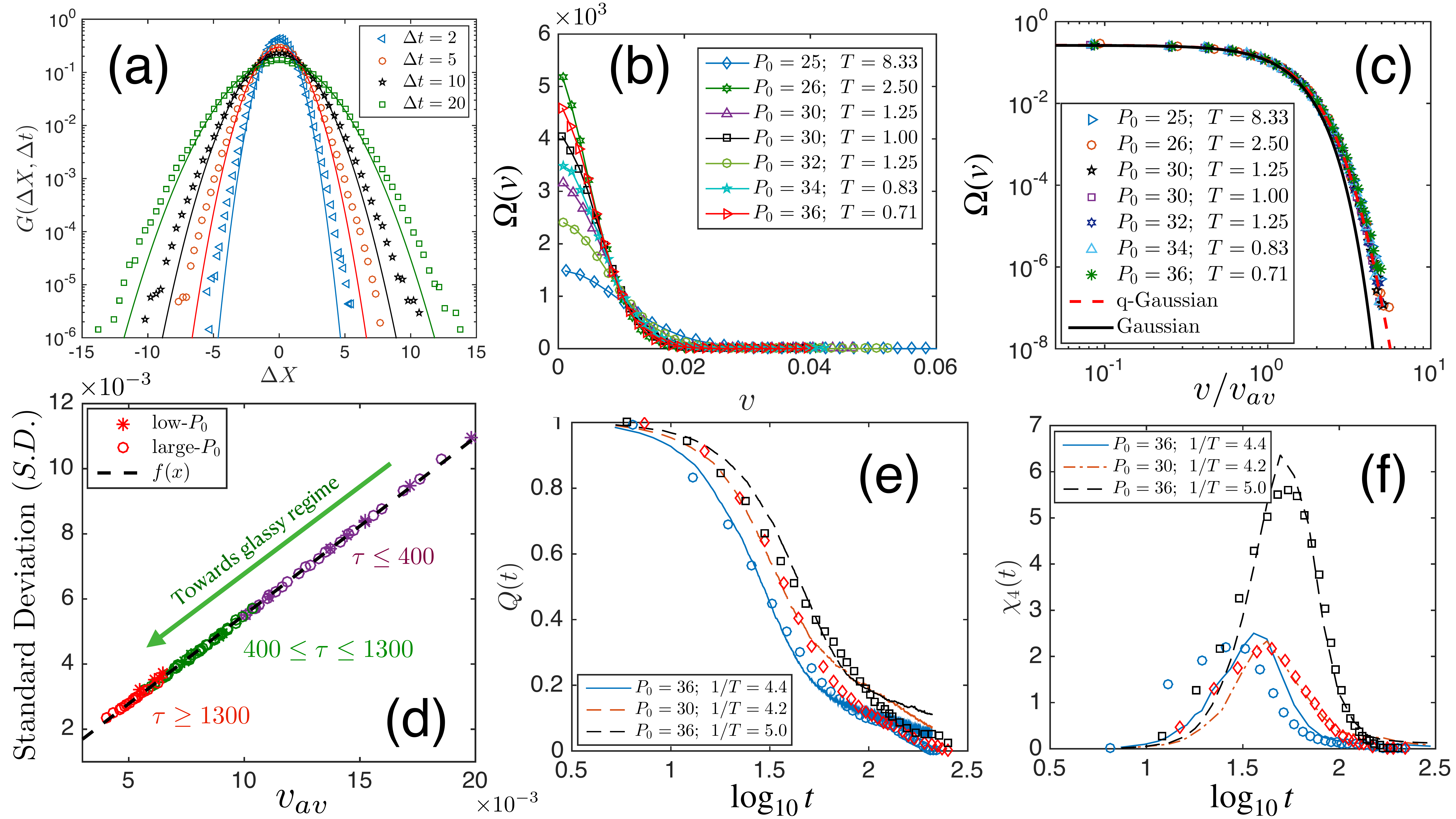}
	\caption{Comparison with experiments. (a) The van-Hove function, Eq. (\ref{vanHovefunction}), for the one-dimensional displacement $\Delta X$ within a time $\Delta t$ in units of $\tau/5$. $P_0=30$ and $T=1.2$ for these results. Symbols are simulation data and lines are fits to Gaussian function. The non-Gaussian nature of $G(\Delta X,\Delta t)$ at large $\Delta X$ are also found in experiments \cite{giavazzi2018,gal2013}. (b) $\Omega(v)$ as a function of $v$ for different $P_0$ and $T$. (c) $\Omega(v)$ for $v/v_{av}$ for the data in (b) follow a master curve that is well-described by $q$-Gaussian, and not Gaussian, distribution. (d) Standard deviation (S.D.) as a function of $v_{av}$ at various $P_0$ and $T$ follows a straight line. We have chosen different colors for $\tau$ in three regimes shown in the figure. Point in the plot corresponding to a system with higher $\tau$ shifts towards the origin. The dotted line is a linear fit to the data: $f(x)=a+bx$ with $a\simeq 5.5\times10^{-5}$ and $b\simeq0.55$. The results in (c) and (d) are in agreement with experiments \cite{lin2020}. (e) $Q(t)$ at different $P_0$ and $T$. The simulation parameters are chosen in such a way that $Q(t)$ qualitatively agree with the experimental data in Ref. \cite{atia2018} (in Fig. S1(e)). Lines are CPM results and symbols are experimental data. (f) Comparison of $\chi_4(t)$ in CPM (lines) corresponding to the same $P_0$ and $T$ as in (e) and experimental data (symbols) from Fig. S1(f) in \cite{atia2018}.}
	\label{exptcompfig}
\end{figure*}

Finally, we compare the stretching exponent $\beta$ \cite{vaibhav2020,xia2001} that describes decay of the overlap function $Q(t)\sim \exp[-(t/\tau)^\beta]$. The RFOT expression for $\beta$ (Appendix \ref{stretchingexpo}) is 
\begin{equation}\label{rfotbeta}
\beta=\mathcal{A}\left[1+\left\{\f{\mathcal{B}[k_1-k_2(P_0-23)]}{T-T_K+\varkappa_c(P_0-23)}\right\}^2\right]^{-1/2},
\end{equation}
where $\mathcal{A}$ and $\mathcal{B}$ are two constants; we fit Eq. (\ref{rfotbeta}) with the simulation data for $P_0=25$, as shown in Fig. \ref{summarycomp}(e), and obtain $\mathcal{A}=0.62$ and $\mathcal{B}=0.3$. We then compare the RFOT predictions with simulation data for different $P_0$ as shown in Fig. \ref{summarycomp}(f) for four different $T$. Again, the trends for $\beta$ agree quite well with theoretical predictions.

\section{Comparison with experiments} 
\label{compexpt}
We now demonstrate the applicability of CPM to biological systems through comparison of theoretical predictions with existing experimental data. Instead of a detailed comparison with a particular system, as is common for biophysical modeling, our aim here is to illustrate that CPM captures the key characteristics of dynamics in a wide class of confluent cellular monolayer. 
An important characteristic of glassiness is the non-Gaussian nature of the van-Hove function, $G(\Delta X,\Delta t)$, which is the probability distribution of displacements, within time $\Delta t$, of the constituent cells in the system, defined as
\begin{equation}\label{vanHovefunction}
G(\Delta X,\Delta t)=\langle \delta(\Delta X- [X_{cm}^\sigma(t_0+\Delta t)-X_{cm}^\sigma(t_0)])\rangle_{\sigma,t_0}
\end{equation}
where the averaging is over all cells and $t_0$. $G(\Delta X,\Delta t)$ is Gaussian at small $\Delta X$ and deviates from the Gaussian behavior at large $\Delta X$ as shown in Fig. \ref{exptcompfig}(a). Similar non-Gaussian behavior of the van-Hove function at large displacement has been reported for the dynamics in a confluent cellular monolayer of MDCK (Madin-Derby Canine Kidney) cells in Ref. \cite{giavazzi2018} and for breast cancer cells in Ref. \cite{gal2013}.

We next compare our simulation results for probability distribution function (PDF) of cell velocities, $v$, with the experiments of Ref. \cite{lin2020}. Tail of the PDF is important as rare events are crucial in glassy dynamics.
Following Ref. \cite{lin2020}, we call the PDF of $v$ as $\Omega(v)$, the circumferentially averaged-PDF, which is defined in Appendix \ref{capdf}. 
Figure \ref{exptcompfig}(b) shows the $\Omega(v)$ for $v$ at different $P_0$ and $T$ in our simulation and the $\Omega(v)$ for the scaled velocity, $v/v_{av}$, where $v_{av}$ is the averaged velocity, follow a master curve (Fig. \ref{exptcompfig}c). We find that $\Omega(v)$ deviates from a Gaussian distribution (Fig. \ref{exptcompfig}c) and well-described by a $q$-Gaussian distribution, $f_{qG}$, defined in the Appendix \ref{qgaussian}.
This result also highlights the distinctive nature of glassiness in a confluent system from that in particulate systems, where $\Omega(v)$ follows a Gaussian distribution \cite{sepulveda2013}. Further, we find that the standard deviation (S.D.) of the velocity distribution linearly depends on $v_{av}$ at different $P_0$ and $T$ (Fig. \ref{exptcompfig}d). In the simulations, we have used state points both in low-$P_0$ and large-$P_0$ regimes at different $T$ and the points in Fig. \ref{exptcompfig}(d) are marked with three different colors based on $\tau$ at that particular $(P_0,T)$. As the system becomes more glassy, that is $\tau$ increases, the points in the $(v_{av},S.D.)$ plot moves towards the origin. As can be seen in Fig. \ref{exptcompfig}(d), this behavior is similar both in low-$P_0$ and large-$P_0$ regimes. These results, i.e., data collapse of $\Omega(v)$ as a function of $v/v_{av}$ and the master curve being described by a $q$-Gaussian function, linear variation of $S.D.$ with $v_{av}$ and movement of data points in $(v_{av},S.D.)$ plot towards origin as the system becomes more glassy are also found in experiments of Ref. \cite{lin2020}.

Finally, we compare the dynamics within CPM with the experiments of Ref. \cite{atia2018}; in particular, we use the data for $Q(t)$ to obtain the values for the control parameters, and then compare the data for four-point correlation functions, $\chi_4(t)$ (defined in Appendix \ref{simdet}), in Figs. \ref{exptcompfig}(e) and (f) respectively. We have chosen different values of $P_0$ and $T$ to best represent $Q(t)$ presented in Fig. S1(e) in the Sup. Mat. of Ref. \cite{atia2018}, as shown in Fig. \ref{exptcompfig}(e) by symbols and lines represent CPM data where we rescaled and shifted time in the theory.
Note that the time-scale in a Monte-Carlo simulation is arbitrary, therefore, a scaling of this time is not important. We have rescaled the simulation time (in logarithmic scale) by a factor of $2.5$ and shifted it by $\approx 1.9$ to show them on the same scale as the experimental data.
We plot the corresponding $\chi_4(t)$ from our simulation as a function of the same rescaled time as in Fig. \ref{exptcompfig}(e) along with the experimental data in Fig. \ref{exptcompfig}(f). 
Qualitative agreement of $\chi_4(t)$ between CPM results and experimental data \cite{atia2018} in Fig. \ref{exptcompfig}(f) demonstrates that CPM does qualitatively capture the information of dynamic heterogeneity, given by $\chi_4(t)$ \cite{smarajit2009}, in the experimental system.

\section{Discussion and conclusion} 
\label{disc}
Complete confluency imposes a strong geometric restriction bringing about two different regimes as $P_0$ is varied.
Our theory traces the unusual sub-Arrhenius behavior to the distinctive nature of interaction potential resulting via the perimeter constraint and shows up in a regime where the system is about to satisfy this constraint. 
Qualitative similarities of the results presented here with those from vertex-based simulations \cite{bi2015,sussman2018,li2021} suggest glassiness in such systems depends on two key elements, first, the energy function, and second, the confluent nature, and {\em not} the microscopic details, of the models. 
We believe, the RFOT theory that we have developed is applicable to a general confluent system and not restricted to CPM. In particular, the simulation results of vertex-based models can be understood within the RFOT theory that we have developed here \cite{manoj2020}.
The three predictions of the theory that we have discussed, namely super-Arrhenius behavior in a different region of low-$P_0$ regime, super-Arrhenius and constant fragility in a model with $\lambda_P=0$ and the stretching exponents at different $P_0$ agree well with our simulation data within CPM. These predictions can be easily tested in vertex-based simulations, such results will further establish the similarity (or the lack of it) of such models with CPM.

Vertex-model simulations have argued the rigidity transition controls the glassy dynamics, and the observed shape index, $q$, has been interpreted as a structural order parameter for glass transition \cite{park2015,bi2015,bi2016}. Our study shows that these results are {\em not} generic for confluent systems. 
The rigidity transition in vertex-models as geometric incompatibility in the two regimes have been studied in the literature \cite{moshe2018,merkel2018}; our results, however, seem to indicate this transition is a result of the nonequilibrium nature of the $T1$ transitions within Vertex model.
The lowest value of $q$ is determined by geometric restriction in the low-$P_0$ regime whereas it is proportional to $P_0$ in the large-$P_0$ regime although glassiness is found in both; thus, $q$ can not be treated as an order parameter for glassy dynamics within CPM.

Control parameters of glassiness in a confluent system are different from those in particulate systems. The experiments of Ref. \cite{malinverno2017} on human mammary epithelial MCF-10A cells show that expression of RAB5A, that does not affect number density, fluidizes the system. Careful measurements reveal RAB5A affects the junction proteins in cortex that determines the target perimeter $P_0$ \cite{malinverno2017,palamidessi2019}, which is a control parameter for glassiness in such systems. 
We emphasize that the presence of lattice in CPM only affects the quantitative values of the parameters: for example, on a square lattice the minimum perimeter configuration for a certain area is a square. However, this, we believe, does not affect the qualitative behaviors and the physics behind them.

Apart from biological importance for simulating confluent systems, CPM provides an interesting system to study from purely theoretical point of view to understand glassy dynamics in a new light; the well-defined equilibrium limit and discrete nature of the model are advantages over vertex models. It is important to understand the source of the sub-Arrhenius nature of relaxations in more detail, though it is unusual in particulate system, it is not unique to confluent systems. Is it possible to define models of point particles with specific interaction potential to find similar behavior?

We have demonstrated that simulation results of CPM agree well with existing experimental data on diverse confluent cellular systems: the non-Gaussian van-Hove function \cite{giavazzi2018,gal2013}, the nontrivial velocity distribution \cite{lin2020}, relation between the standard deviation of velocities with their averages \cite{lin2020}, the behavior of two- and four-point functions \cite{atia2018}, limiting value of observed shape index in the low-$P_0$ regime \cite{park2015}, etc have also been found in experiments. 
The non-Gaussian velocity distribution highlights the distinctive nature of glassiness in confluent systems compared to that in particulate systems \cite{lin2020,sepulveda2013}; agreement of this distribution between CPM and experiments on a variety of systems is, therefore, encouraging. A crucial result of our simulations is the presence of glassy behavior in the large-$P_0$ regime, where vertex-model simulations suggest absence of glassiness \cite{bi2015,bi2016}; the difference seems to come from the details of how $T1$ transitions are included within the two models. 
The complex biochemical reactions that governs dynamics in a biological system is represented by $T$ in CPM. As metabolic activity reduces, self-propulsion, which is absent in the current model, as well as $T$ also decrease. Therefore, experimental verification of presence or absence of glassiness in the large-$P_0$ regime as metabolism is decreased in a biological system along the lines of Refs. \cite{palamidessi2019,decamp2020} can be a critical test for applicability of CPM.

\section{Acknowledgements}
We thank Mustansir Barma and Chandan Dasgupta for many important and enlightening discussions and critical comments on the manuscript. We also thank Tamal Das, Kabir Ramola, Navdeep Rana, Kallol Paul and Pankaj Popli for discussions and Cristina Marchetti for comments on the manuscript. We acknowledge support of the Department of Atomic Energy, Government of India, under Project Identification No. RTI 4007

\appendix
\section{Simulation details} 
\label{simdet}
For the results presented here, unless otherwise specified, we use a system of size $120\times120$ with 360 cells and an average cell area of 40. The minimum possible perimeter for a cell with area 40 on a square lattice is 26. We start with a rectangular cell initialization with $5\times 8$ sites having same Potts variable and equilibrate the system for at least $8\times 10^5$ MC time steps before collecting data. We have set $\lambda_A=1$ and $\lambda_P=0.5$ for the results presented here. We have checked that the behavior remains same for other values of $\lambda_P$ as well as  cell sizes (data not presented).

{\bf Mean square displacement and self-overlap function:} Dynamics is quantified through the mean square displacement ($MSD$) and the self-overlap function, $Q(t)$. $MSD$ is defined as 
\begin{align}
MSD=\overline{\f{1}{N}\sum_{\sigma=1}^N \langle(\mathbf{X}_{cm}^\sigma(t+t_0)-\mathbf{X}_{cm}^\sigma(t_0))^2\rangle_{t_0}},
\end{align}
where $\mathbf{X}_{cm}^\sigma(t)$ is center of mass of cell $\sigma$ at time $t$, $\langle \ldots\rangle_{t_0}$ denotes averaging over initial times $t_0$ and the overline implies an averaging over ensembles. Unless otherwise stated, we have taken 50 $t_0$ averaging and 20 configurations for ensemble averaging. $Q(t)$ and $\chi_4(t)$ are defined as
\begin{align}\label{qandchi4def}
	Q(t)&=\overline{\f{1}{N}\sum_{\sigma=1}^N \langle W(a-|\mathbf{X}_{cm}^\sigma(t+t_0)-\mathbf{X}_{cm}^\sigma(t_0)|)\rangle_{t_0}} \nonumber\\
	&=\overline{\langle\tilde{Q}(t)\rangle_{t_0}}, \nonumber \\
	\chi_4(t)&=N\overline{( \langle\tilde{Q}(t)^2\rangle_{t_0}-\langle\tilde{Q}(t)\rangle_{t_0}^2)}
\end{align}
where $W(x)$ is a heaviside step function
\begin{align}
W(x)=\begin{cases}
1  & \text{if } x\geq 0\\
0 & \text{if } x<0
\end{cases}
\end{align}
and $a$ is a parameter that we set to 1.12. 

\section{Stretching exponent for the decay of the self-overlap function}
\label{stretchingexpo}
It is well-known that the decay of self-overlap function, $Q(t)$, in a glassy system can be described through a stretched exponential function \cite{vaibhav2020}, the Kohlrausch-Williams-Watts (KWW) formula \cite{kohlrausch1854,williams1970} given by,
\begin{equation}
Q(t) = A  \exp[{-(t/\tau)^\beta}],
\label{eq :stexpSM}
\end{equation}
where $A$ is a constant, of the order of unity, $\tau$, the relaxation time and $\beta$ is the stretching exponent. RFOT theory allows calculation of $\beta$ through the fluctuation of local free energy barriers $\Delta F$ \cite{xia2001}. We assume that $\Delta F$ follows a Gaussian distribution given by,
\begin{equation}
P(\Delta F) = \frac{1}{\sqrt{2\pi\sigma_F^2}}\exp\left[{-\frac{(\Delta F - \Delta  F_0)^2}{2\sigma_F^2}}\right]
\label{eq :distributionSM}
\end{equation}
where $\Delta  F_0$ is the mean of the distribution and $\sigma_F$ is the standard deviation, which gives a measure of the fluctuation. Following Xia and Wolynes \cite{xia2001}, we obtain $\beta$ as
\begin{equation}
\beta = \Big [ 1 + \big(\frac{\sigma_F}{T}\big)^2\Big]^{-\frac{1}{2}},
\label{eq: betaSM}
\end{equation}
where we have set Boltzmann constant $k_B$ to unity.
For the Gaussian distribution of $\Delta F$, we obtain \cite{xia2001},
\begin{equation}\label{delscavscSM}
\frac{\delta s_c}{\langle s_c \rangle} \sim \frac{\sigma_F}{\Delta F_0},
\end{equation}
with $\delta s_c \sim \sqrt{\Delta  C_p/V}$, where $V\sim \xi^d$ is the typical volume of the mosaics. In the low-$P_0$ regime, where we have compared our RFOT theory predictions with the simulation results, the length scale $\xi$ of the mosaics, Eq. (\ref{xieqSM1}), is given by,
\begin{equation}
\xi  \sim \Big [ \frac{k_1-k_2(P_0-P_0^{ref})}{T -T_K+\varkappa_c (P_0-P_0^{ref})} \Big ]^{1/(d-\theta)}
\label{eq: lengthscaleSM}
\end{equation}
\begin{equation}\label{scavSM}
\text{and, } \langle s_c \rangle \sim \frac{\Delta C_p}{T_K}[T -T_K+\varkappa_c (P_0-P_0^{ref})].
\end{equation}
Using Eqs. (\ref{eq: lengthscaleSM}) and (\ref{scavSM}), we obtain
\begin{equation}
\frac{\delta s_c}{\langle s_c \rangle} \propto [k_1-k_2(P_0-P_0^{ref})]^{-1}.
\label{eq: ratioSM}
\end{equation}
The mean free energy barrier $(\Delta F_0)$  is obtained, by using $R=\xi$ in Eq. (\ref{rfotenergySM}), as
\begin{equation}
\Delta F_0 \propto \Big [ \frac{T[k_1-k_2 (P_0-P_0^{ref})]^2}{T -T_K+\varkappa_c (P_0-P_0^{ref})} \Big ].
\label{eq: F0SM}
\end{equation}
Using Eqs. (\ref{eq: ratioSM}), (\ref{eq: F0SM}) and (\ref{delscavscSM}) in Eq. (\ref{eq: betaSM}), we obtain $\beta$ as
\begin{equation}\label{betalowP01SM}
\beta  = \Big [ 1 + \Big \{ \frac{\mathcal{B}[k_1-k_2(P_0-P_0^{ref})]}{T -T_K +\varkappa_c (P_0-P_0^{ref})} \Big \}^2 \Big]^{-1/2}
\end{equation}
where $\mathcal{B}$ is a constant. It is well-known that RFOT theory predicts the correct trends of $\beta$, but the absolute values differ by a constant factor even for a particulate system \cite{xia2001}. Since we are interested in the trend of $\beta$ as a function of $P_0$, we multiply Eq. (\ref{betalowP01SM}) by a constant $\mathcal{A}$ to account for this discrepancy and obtain
\begin{equation}\label{betalowP02SM}
\beta  = \mathcal{A}\Big [ 1 + \Big \{ \frac{\mathcal{B}[k_1-k_2(P_0-P_0^{ref})]}{T -T_K +\varkappa_c (P_0-P_0^{ref})} \Big \}^2 \Big]^{-1/2}.
\end{equation}
The constants $k_1$, $k_2$, $T_K$ and $\varkappa_c$ are already determined, $\mathcal{A}$ and $\mathcal{B}$ are obtained through the fit of Eq. (\ref{betalowP02SM}) with the simulation data for $P_0=25$ as a function of $T$.

\section{Calculation of $\Omega(v)$ in our simulation}
\label{capdf}
Ref. \cite{lin2020} looks into the circumferentially averaged-PDF  [$\Omega(v)$]. Following Ref. \cite{lin2020}, we have obtained the $\Omega(v)$ as follows: We calculate velocities of different cells from their displacements, $r$, of their centers of mass after 100 MC steps and define $v=r/100$. We then use a velocity-grid labeled by $i$ and obtain the number of velocity events, $\mathcal{O}_i$, within a range $v^{i}$ and $v^{i+1}$. Finally, we obtain 
\begin{equation}
\Omega(v_i)=\frac{1}{N_v}\frac{\mathcal{O}_i}{2\pi v_{i}dv_i}
\end{equation}
where $v_i=(v^{i}+v^{i+1})/2$, $dv_i=(v^{i+1}-v^{i})$ and $N_v$ is the total number of velocity events. 

\section{$q$-Gaussian distribution}
\label{qgaussian}
The $\Omega(v)$ is well described by a $q$-Gaussian distribution, $f_{qG}$, defined as 
\begin{equation}\label{qgaussianSM}
f_{qG}(v)=A_q(1+B_qv^2)^{-\lambda_q}
\end{equation}
where $\lambda_q=1/(q-1)$, $A_q=(1/\pi)(\lambda_q-1)B_q$, and $B_q=(\pi/4)\left[\Gamma(\lambda_q-3/2)/\Gamma(\lambda_q-1)\right]^2$. $\Gamma(\ldots)$ is the Gamma function. From the fit we obtain $q=1.06$ ($q$ here is different from shape index).

%\bibliography{cpmref.bib}
%\bibliographystyle{apsrev4-1}

%merlin.mbs apsrev4-1.bst 2010-07-25 4.21a (PWD, AO, DPC) hacked
%Control: key (0)
%Control: author (72) initials jnrlst
%Control: editor formatted (1) identically to author
%Control: production of article title (-1) disabled
%Control: page (0) single
%Control: year (1) truncated
%Control: production of eprint (0) enabled
%

\end{document}